\documentclass[sigconf]{acmart}

\usepackage{amsmath,amsfonts}
\usepackage{graphicx}
\usepackage[table]{xcolor}
\usepackage[nameinlink]{cleveref}
\usepackage{siunitx}
\usepackage{xspace}
\usepackage{multirow}
\usepackage{enumitem}
\usepackage{makecell}
\usepackage{soul}
\usepackage[linesnumbered,ruled]{algorithm2e}

\newcommand{\ourtool}{\textsc{MuoFuzz}\xspace}
\newcommand{\afl}{\textsc{AFL}\xspace}
\newcommand{\aflpp}{\textsc{AFL++}\xspace}
\newcommand{\mopt}{\textsc{MOPT}\xspace}

\sethlcolor{green}
\crefname{algocf}{algorithm}{algorithms}
\Crefname{algocf}{Algorithm}{Algorithms}

\colorlet{myblue}{cyan!60}
\colorlet{myblueopaque}{cyan!20}

\usepackage{tikz} 
\usetikzlibrary{fit,calc}
\newcommand*{\tikzmk}[1]{\tikz[remember picture,overlay,] \node (#1) {};\ignorespaces}
\newcommand{\boxit}[1]{\tikz[remember picture,overlay]{\node[yshift=1pt,fill=#1,opacity=.33,fit={(A)($(B)+(.83\linewidth,.8\baselineskip)$)}] {};}\ignorespaces}

\AtBeginDocument{%
  }

\acmConference[ICSE 2026]{48th IEEE/ACM International Conference on Software Engineering}{April 12--18,
  2026}{Rio de Janeiro, Brazil}
\acmISBN{XXX-X-XXXX-XXXX-X/XX/XX}

\begin{document}

\title{On Interaction Effects in Greybox Fuzzing}

\author{Konstantinos Kitsios}
\email{konstantinos.kitsios@uzh.ch}
\affiliation{
  \institution{University of Zurich}
  \city{Zurich}
  \country{Switzerland}
}

\author{Marcel B{\"o}hme}
\email{marcel.boehme@acm.org}
\affiliation{
  \institution{Max Planck Institute for Security and Privacy}
  \city{Bochum}
  \country{Germany}
}

\author{Alberto Bacchelli}
\affiliation{
  \institution{University of Zurich}
  \city{Zurich}
  \country{Switzerland}
}
\email{bacchelli@ifi.uzh.ch}

\begin{abstract}
A greybox fuzzer is an automated software testing tool that generates new test inputs by applying randomly chosen mutators (e.g., flipping a bit or deleting a block of bytes) to a seed input in random order and adds all coverage-increasing inputs to the corpus of seeds. We hypothesize that the \emph{order} in which mutators are applied to a seed input has an impact on the effectiveness of greybox fuzzers.
In our experiments, we fit a linear model to a dataset that contains the effectiveness of all possible mutator pairs and indeed observe the conjectured interaction effect. This points us to more efficient fuzzing by choosing the most promising mutator sequence with a higher likelihood.

We propose \textsc{MuoFuzz}, a greybox fuzzer that learns and chooses the most promising mutator sequences. \textsc{MuoFuzz} learns the conditional probability that the next mutator will yield an interesting input, given the previously selected mutator.
Then, it samples from the learned probability using a random walk to generate mutator sequences.
We compare the performance of \textsc{MuoFuzz} to \aflpp, which uses a fixed selection probability, and \textsc{MOPT}, which optimizes the selection probability of each mutator in isolation.
Experimental results on the FuzzBench and MAGMA benchmarks show that \textsc{MuoFuzz} achieves the highest code coverage and finds four bugs missed by \textsc{AFL++} and one missed by both \textsc{AFL++} and \textsc{MOPT}.
\noindent \textit{Data and material: \href{https://doi.org/10.5281/zenodo.17391100}{https://doi.org/10.5281/zenodo.17391100}}
\end{abstract}

\ccsdesc[500]{Security and privacy~Software security engineering}
\ccsdesc[500]{Software and its engineering~Software testing and debugging}

\keywords{software security, fuzzing, mutation strategy}


\maketitle

\section{Introduction}\label{sec:introduction}

Mutation-based greybox fuzzing aims to automatically detect unexpected software behaviour~\cite{bohme2016markov,bohme2017directed,klees2018evaluating,fuzzingbook,cerebro,dodrio}.
Given a target program and a corpus of seed inputs (or simply \emph{seeds}) to that program, fuzzers select one seed and mutate it to produce a mutated input (or simply \emph{input}), which they feed to the target program and gather code coverage feedback.
If the input covers new code, it is added to the seed corpus for further mutations.
By repeating this process for millions of inputs, fuzzers automatically reach and test deep program locations.

\afl~\cite{afl} and its more recent successor, \aflpp~\cite{afl++}, are the most widely used fuzzers.
To mutate a seed, \aflpp relies on \num{32} mutators, ranging from simple random bit flipping to disruptive deletions of byte blocks from the seed.
Many of these mutators are applied consecutively to the seed (forming a \textit{mutator sequence}) and the algorithm that controls how mutator sequences are generated is called \emph{mutation strategy}.
The mutation strategy of \aflpp is straightforward:
each of the \num{32} mutators has a fixed probability of being selected.
Researchers~\cite{mopt,seamfuzz,slopt,havocMAB} have proposed mutation strategies that adjust the probability of each mutator, such that mutators that produce more coverage-increasing inputs (also called \emph{interesting} inputs) have a higher probability of being selected.

The aforementioned studies model and leverage the probability that each mutator will yield an interesting input in isolation, and then sample $l$ mutators from this probability to create a mutator sequence of length $l$.
This process assumes that each mutator in the sequence is independent of the others: the probability of selecting the next mutator does not take into account the already chosen mutators.

We hypothesize that there may exist an interaction effect between mutators, which can be leveraged to further increase the number of interesting inputs and, in turn, the fuzzer's performance.
Based on this hypothesis, we propose a threefold contribution:
\begin{itemize}[leftmargin=1em]
    \item Empirical evidence that some mutators, when combined, produce more interesting inputs than others.
    \item A mutation strategy, implemented into \ourtool~\cite{ours}, that leverages our first finding by sampling the next mutator from a probability distribution conditioned on the previously selected mutator.
    \item Empirical evidence on the effectiveness of \ourtool by comparing against state-of-the-art fuzzers in terms of achieved code coverage and found bugs.
\end{itemize}

For our first contribution, we fit a linear model where the two independent variables are the two mutators in a sequence of length two and the dependent variable is the number of interesting inputs produced by this sequence.
We collect our dataset by running AFL++ in \num{13} target programs.
By running an Analysis of Variance (ANOVA)~\cite{anova} on the fitted model, we find that the interaction term explains a statistically significant proportion of the variance. This means that the interaction effect between two mutators affects the number of interesting inputs these mutators produce.

For our second contribution, we leverage this newly found interaction effect: We propose a strategy for generating mutator sequences where the probability of
selecting the next mutator is conditioned on the previously selected mutator.
We implement this strategy into \ourtool (\textbf{Mu}t + Du\textbf{o} + \textbf{Fuzz}), which works in two phases:
During the (1) \emph{training phase}, \ourtool learns the conditional probability that a mutator will yield an interesting input given the previous mutator in the sequence;
during the (2) \emph{guided mutation phase}, \ourtool samples from the learned probability using a random walk~\cite{markov}.

For our final contribution, we compare the performance of \ourtool with \aflpp, which uses a fixed selection probability, and \mopt~\cite{mopt}, which optimizes the selection probability of each mutator in isolation, without considering the interaction effect between mutators.
\ourtool achieves the highest code coverage in \num{10} out of \num{13} FuzzBench~\cite{fuzzbench} programs. 
Moreover, it finds four bugs that \aflpp missed, as well as one bug that both \aflpp and \mopt missed in the MAGMA benchmark~\cite{magma}.

\section{Background}\label{sec:background}
In this section, we present the inner workings of mutation-based greybox fuzzers to provide the necessary background for the rest of the paper.


\begin{algorithm}
\scriptsize
\caption{\small Mutation-based greybox fuzzing. The part we modify is \colorbox{myblueopaque}{highlighted}.}
\label{alg:greybox_fuzzing}
\SetKwInOut{Input}{Input}
\SetKwInOut{Output}{Output}
\Input{Target program $p$, seed corpus $S$, set of mutators $\mathcal{M}$}
\Output{Corpus with crashing inputs $Crash$}
    $Crash \gets \emptyset$\;
    \Repeat{timeout reached}{
        $s \gets SelectSeed(S)$\;
        \For{$i_{input} = 1$ \KwTo $num\_inputs\_for\_this\_seed$}{ 
            \tikzmk{A}
            $M \gets \emptyset$\;
            \For{$n = 1$ \KwTo $l$}{
                $m_n \sim \Pr(\mathcal{M});$ \tcp{fixed for AFL++; learned for MOPT; we propose $\Pr(\mathcal{M} \, \mid \, m_{n-1})$}
                $M \gets M \cup \{m_n\}$\;
            }
            \tikzmk{B}
            \boxit{myblue}
            $s' \gets Mutate(s, M)$\;
            $Crash, S \gets Execute(p, s', Crash, S)$\;
        }
    }
    \Return{Crash}\;
\end{algorithm}

\subsection{Mutation-based Greybox Fuzzing}\label{ssec:mutation_based_greybox_fuzzing}
A mutation-based greybox fuzzer takes as input a target program and a corpus of seed inputs.
These initial seeds are usually human-written inputs aiming to provide a good starting point for mutation.
The fuzzer automatically generates inputs for the target program by following the steps of \Cref{alg:greybox_fuzzing}.
First, it selects a seed from the corpus to mutate (line 3).
The probability of selecting a seed depends on heuristic rules:
For example, seeds that have been selected in the past have a lower probability, while seeds that are smaller in size have a higher probability. Then, the fuzzer decides the number of mutated inputs that will be produced from this seed (loop limit in line 4).
This number ranges from \num{16} to many thousands and depends on similar heuristics.
Afterward, the fuzzer mutates the seed to generate an input, following the mutation strategy described in the next section (\Cref{ssec:aflpp_mut_strategy}).
Finally, the fuzzer feeds the input to the program (line 11) and monitors its behaviour:
If the input crashes the program, then it is considered a \emph{potential} bug and is returned for human inspection (line 14).
If the input achieves new code coverage, it is considered \textit{interesting} and is added to the seed corpus to be further mutated in future iterations~\cite{fuzzingbook}.

\begin{table}
\centering
\scriptsize
\caption{\aflpp mutators and their selection probability.}
\begin{tabular}{lp{5.3cm}ll}
\textbf{ID} & \textbf{Description} & \textbf{Type}  & \textbf{Probability} \\ \midrule
1 & flip a random bit & unit & \rule{7.9mm}{1mm} \scriptsize 0.043\\
2 & replace a random byte with an interesting value & unit & \rule{6.4mm}{1mm} \scriptsize 0.035\\
3 & replace two adjacent bytes with interesting values & unit & \rule{4.3mm}{1mm} \scriptsize 0.023\\
4 & replace two adjacent bytes with interesting values (be*) & unit & \rule{4.3mm}{1mm} \\
5 & replace four adjacent bytes with interesting values & unit & \rule{4.3mm}{1mm} \\
6 & replace four adjacent bytes with interesting values (be) & unit & \rule{4.3mm}{1mm} \\
7 & subtract a value between 1 and 35 from a random byte & unit & \rule{6.4mm}{1mm} \\
8 & add a value between 1 and 35 to a random byte & unit & \rule{7.1mm}{1mm} \\
9 & subtract a value between 1 and 35 from two adjacent bytes & unit & \rule{4.3mm}{1mm} \\
10 & subtract a value between 1 and 35 from two adjacent bytes (be) & unit & \rule{4.3mm}{1mm} \\
11 & add a value between 1 and 35 to two adjacent bytes & unit & \rule{4.3mm}{1mm} \\
12 & add a value between 1 and 35 to two adjacent bytes (be) & unit & \rule{4.3mm}{1mm} \\
13 & subtract a value between 1 and 35 from four adjacent bytes & unit & \rule{4.3mm}{1mm} \\
14 & subtract a value between 1 and 35 from four adjacent bytes (be) & unit & \rule{4.3mm}{1mm} \\
15 & add a value between 1 and 35 to four adjacent bytes & unit & \rule{4.3mm}{1mm} \\
16 & add a value between 1 and 35 to four adjacent bytes (be) & unit & \rule{4.3mm}{1mm} \\
17 & set a random byte to a random value & unit & \rule{6.4mm}{1mm} \\
18 & increase a random byte by 1 & unit & \rule{4.3mm}{1mm} \\
19 & decrease a random byte by 1 & unit & \rule{4.3mm}{1mm} \\
20 & flip all the bits of a random byte & unit & \rule{2.9mm}{1mm} \\
21 & swap a block of bytes between two positions in the seed & chunk & \rule{4.3mm}{1mm} \\
22 & delete a block of bytes & chunk & \rule{6.4mm}{1mm} \\
23 & overwrite a  block of the seed with a dictionary entry & chunk & \rule{7.1mm}{1mm} \\
24 & insert a dictionary entry into a random position of the seed & chunk & \rule{8.6mm}{1mm} \\
25 & overwrite a  block of the seed with an auto-dictionary entry & chunk & \rule{6.4mm}{1mm} \\
26 & insert an auto-dictionary entry into a random position & chunk & \rule{7.9mm}{1mm} \\
27 & select a block from another seed of the corpus, and use it to overwrite a block of the seed & chunk & \rule{9.3mm}{1mm} \\
28 & select a block from another seed of the corpus and insert it into a random position of the seed & chunk & \rule{10mm}{1mm} \\
29 & select a block and insert a copy of it at a different position & chunk & \rule{10mm}{1mm} \\
30 & insert a block of constant bytes to a random position in the seed. The constant block can either be a random value or a part of the original seed. & chunk & \rule{5mm}{1mm} \\
31 & select a block and overwrite another block with it & chunk & \rule{7.1mm}{1mm} \scriptsize 0.039\\
32 & select a block and overwrite it with a fixed byte value, which can be either a random byte or a byte from the original seed & chunk & \rule{3.6mm}{1mm} \scriptsize 0.020 \\
\midrule
\multicolumn{4}{r}{*be = big endian} \\
\end{tabular}
\label{tab:mutators}
\end{table}

\subsection{Mutation Strategy}\label{ssec:aflpp_mut_strategy}
The mutation strategy of AFL++, on top of which we develop our fuzzer, consists of three stages. 
The selected seed is first propagated through the \emph{deterministic stage} where a set of deterministic mutations are applied. For example, all the bits of the seed are flipped, one at a time.
The deterministic stage targets ``low-hanging fruits'' and is not effective in deep program locations due to its simplicity.

Then, the seed goes through the \emph{havoc stage}, which is the most effective of the three stages~\cite{havocMAB}.
The havoc stage comes with \num{32} predefined mutators shown in \Cref{tab:mutators}.
We categorize the mutators into \emph{unit} and \emph{chunk} mutators, following previous work~\cite{havocMAB}.
Unit mutators perform lightweight modifications to the seed, such as flipping a random bit.
Chunk mutators, on the other hand, disruptively modify the seed, for example by deleting a whole block.
AFL++ performs the following steps in the havoc stage.
First, it selects the mutator sequence length $l$ (loop limit in line 6), i.e., how many stacked mutators to apply to the seed.
This number ranges from \num{2} to \num{16}, with lower values having higher probability.
Then, it samples from the set of mutators $l$ times using the predefined probabilities of \Cref{tab:mutators}, creating a sequence of $l$ mutators (lines 7--8).
The mutator sequence is then applied to the seed sequentially to produce the mutated input (line 10).
The input is finally fed into the target program and coverage feedback is collected as described in \Cref{ssec:mutation_based_greybox_fuzzing}.

The third stage of the AFL++ mutation strategy is the \emph{splice stage}.
It disruptively mutates the seed by selecting a random block and concatenating it with a random block from a different seed. 
Both the splice stage and the deterministic stage have less impact than the havoc stage~\cite{havocMAB}, thus this work, similar to previous works~\cite{seamfuzz}, focuses on the havoc stage. For the rest of this paper, the term mutation strategy will refer to the havoc stage of the mutation strategy.

\section{Related Work}\label{sec:related_work}
We present mutation strategies proposed by previous studies to improve the default mutation strategy of \aflpp, which is program-agnostic:
it assigns a fixed, predefined probability to each of the \num{32} mutators, regardless of the target program it tests.
This approach is suboptimal since some mutators could work better in some programs and worse in others.
This led researchers to investigate mutation strategies that adjust the probability of each mutator based on how many interesting inputs the mutator produces.

MOPT~\cite{mopt} is one of the most successful program-adaptive mutation strategies~\cite{sok} and is included in the official \aflpp repository.
It continuously adjusts the mutator probabilities using a genetic algorithm inspired by the Particle Swarm Optimization technique, which aims to find optimal selection probabilities for the \num{32} mutators so that they maximize the likelihood of generating an interesting input.
SLOPT~\cite{slopt} sets a constraint on the mutation strategy of \aflpp: only a single mutator will be used to create a mutator sequence. 
This simplification reduces the problem to the subproblems of (1) what mutator will be used in each sequence and (2) how many times to stack the mutator.
SLOPT views these two problems as multi-armed bandit (MAB) problems and uses the UCB algorithm~\cite{ucb} to solve them.
HAVOC\textsubscript{MAB}~\cite{havocMAB} treats the problems of selecting the optimal length of the mutator sequence $l$, as well as each mutator in the sequence, as two MAB problems, and uses established MAB algorithms to solve them.
SeamFuzz~\cite{seamfuzz} proposes a seed-adaptive mutation strategy, where different mutator selection probabilities are used for different seeds.
Seeds are clustered according to their similarity, and one selection probability is learned for each cluster.
CMFuzz~\cite{cmfuzz} encodes the seeds as context and uses contextual MAB algorithms to solve it, leading to different mutator probabilities for different seeds.
VUzzer~\cite{vuzzer} has two mutators and selects them with a fixed probability, while focusing on optimizing the location of the seeds to mutate.
MendelFuzz~\cite{mendelfuzz} evaluates and improves the deterministic mutation stage, which involves applying simple mutators (e.g., byte flipping) in all the possible seed positions.
Finally, HonggFuzz~\cite{honggfuzz}, similar to AFL++, implements a program-agnostic mutation strategy with slightly different mutators than AFL++.

All of these studies generate mutator sequences by considering each mutator in the sequence as independent from the others.
This assumption allows the use of MAB algorithms, where, by definition, each arm is independent from the others~\cite{mab}.
Our work focuses on selecting the optimal mutator conditioned on the previously selected one, which makes it orthogonal to previous work.

\section{Investigating The Interaction Effect Between Mutators}\label{sec:rq1}
We hypothesize that some mutators, when combined, are more effective than others and we  experimentally test our hypothesis.

\vspace{10pt}
{
\setlength{\fboxsep}{7pt}

\noindent
\fcolorbox{black}{gray!8}{
  \parbox{0.44\textwidth}{
    \textbf{RQ1.} Can we measure an interaction effect between two mutators in a mutator sequence?
  }
}
}
\vspace{10pt}

To test the existence of the interaction effect, we collect a dataset that captures the effectiveness (i.e., number of interesting inputs discovered) of each possible pair of mutators. 
We note that other factors, like mutator position, can contribute to the number of interesting inputs discovered by a mutator pair.
However, since fuzzers perform hundreds of millions of mutations, it is reasonable to assume that these factors even out across all mutator pairs.

Then, we fit a linear model with an interaction term to the collected dataset and conduct a two-way ANOVA~\cite{anova} to test the significance of the model's interaction term. 
To collect the dataset, we run \aflpp in 13 target programs following previous work~\cite{mopt, bohme2016markov, seamfuzz, darwin, havocMAB}, as it is the most widely used fuzzer.
However, our methodology is applicable to sets of mutators other than \aflpp, and we expect our findings to generalize to other mutation-based greybox fuzzers like libFuzzer~\cite{libFuzzer} that have a similar set of mutators.
The details of the dataset collection process are provided in \Cref{ssec:motiv_dataset_collection}, followed by the model fitting in \Cref{ssec:motiv_model_fitting} and the results in \Cref{ssec:motiv_results}.

\subsection{Dataset Collection}\label{ssec:motiv_dataset_collection}

\begin{figure}
  \centering
  \includegraphics[width=0.5\textwidth]{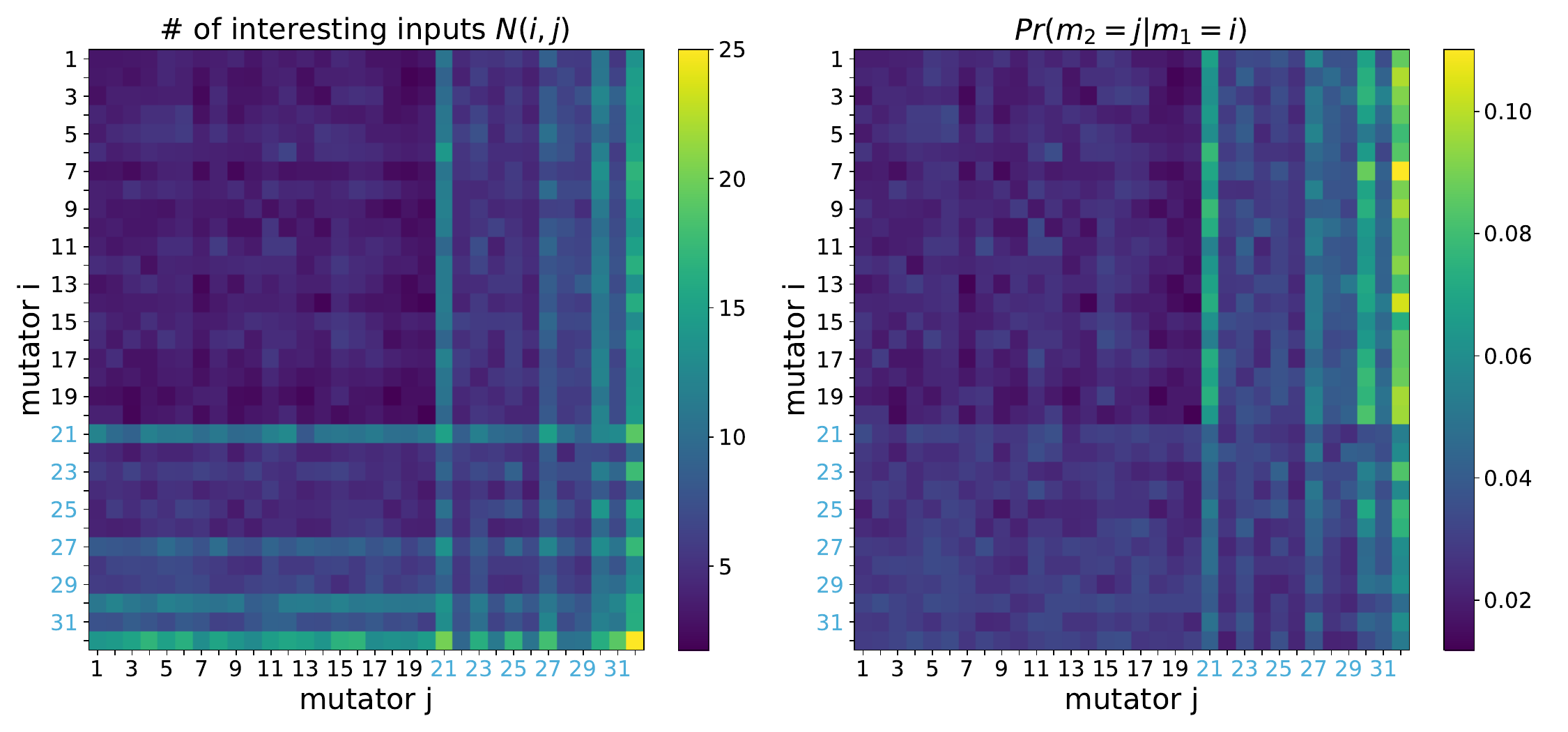}
  \caption{
  The left figure shows the number of interesting inputs generated by the mutator sequence $\langle i,j \rangle$, where $i$ is the mutator in the row and $j$ is the mutator in the column.
  The conditional probability on the right figure is derived by dividing each row of 
  the left figure by its sum.
  As such, each row $i$ of the right matrix is a probability distribution, denoting how probable is that selecting $j$ as the second mutator will yield an interesting input, given that the first mutator is $i$. These concern the \texttt{freetype} target program.}
  \Description{Learned probability matrices  for the freetype program.}
  \label{fig:matrices}
\end{figure}

Let $\mathcal{M}=\{1, 2, ..., M\}$ be the IDs of the available mutators and let $\langle m_1, m_2, ..., m_l \rangle \in \mathcal{M}^l$ denote a mutator sequence of length $l$.
To answer our first research question, we need a dataset containing the number of interesting inputs generated by each possible pair of mutators $\langle i,j \rangle \in \mathcal{M}^2$.
To collect this dataset, we first select 13 target programs that were used in previous work~\cite{seamfuzz,dipri}.
The programs are part of the FuzzBench~\cite{fuzzbench} benchmark and are shown in the first column of \Cref{tab:results_rq2}.
More details on their selection are provided in \Cref{ssec:evaluation_rq2}.
We study each program individually, since some mutator sequences may be efficient in some programs but not in others.
Our goal is to generate data points $N(i,j), i,j \in \mathcal{M}$ for each program that contain the number of interesting inputs generated by the mutator sequence $\langle i,j \rangle$.
To do this, we fuzz the selected target programs using AFL++ with three modifications:

\begin{description}[leftmargin=1em]
    \item[\textbf{M1}:] We limit the length of the mutator sequences to $l=2$ to study the interaction effect between $2$ mutators.
    Studying the interaction effect of $l>2$ mutators is computationally challenging, as we show in \Cref{sssec:more_mutators}, where we study the interaction effect of triplets by using $l=3$.
    We observe that, due to combinatorial explosion, most triplets remain unobserved during training because they do not produce a coverage-increasing input.
    
    \item[\textbf{M2}:] We sample the mutators uniformly at random instead of following the default probabilities of \aflpp, to gather the same amount of data for each mutator pair.
    
    \item[\textbf{M3}:] We maintain a $|\mathcal{M}| \times |\mathcal{M}|$ matrix $N(i, j)$ that holds the number of interesting inputs produced by the sequence $\langle i,j \rangle$.
    Each time $\langle i,j \rangle$ generates an interesting input, we increase $N(i, j)$ by $1$.

\end{description}

We fuzz each target program with \aflpp for \num{20} trials, hence we obtain one matrix for each trial $k \in \{1, 2, ..., 20\}$,  which we call $N(i, j)^{(k)}$.
We experiment with different values for the length $T$ of the fuzzing campaigns, namely $T \in \{1, 5, 24\}$ hours. 
We find that the significance of the interaction effect does not change across the three values of $T$
because \aflpp discovers most of the interesting inputs within the first hours of fuzzing.
\Cref{fig:matrices} (left) shows the collected data $N(i, j)$ averaged over the trials $k$ for the \texttt{freetype} program, while the plots for the rest of the programs are available in our replication package~\cite{ours}.

\subsection{Model Fitting}\label{ssec:motiv_model_fitting}
We use the dataset $N(i, j)^{(k)}$ to train a linear model that, given two mutators $i,j \in \mathcal{M}$, will predict the expected number of interesting inputs generated by the sequence $\langle i,j \rangle$ in a fuzzing campaign for the given target program.
We selected a linear model for simplicity and interpretability. Given the high goodness-of-fit in \Cref{tab:results_rq1}, we did not try higher-order models.
The linear model has categorical independent variables (both $i$ and $j$ are categorical) and a numerical dependent variable~\cite{DraperSmith1998}, so it is of the form:
\begin{equation}\label{eq:linear_model}
    \hat{N}(i,j) = \mu + \alpha_i + \beta_j + \gamma_{ij}
\end{equation}
where $\mu$ is the constant bias, $\alpha_i$ is the effect of the first mutator in the sequence, $\beta_j$ is the effect of the second mutator in the sequence, and $\gamma_{ij}$ is the interaction effect between the two mutators.

\begin{table}
\centering
\caption{The linear model with the interaction term ($\gamma_{ij} \neq 0$) has higher goodness-of-fit than the model without the interaction term ($\gamma_{ij}=0$).}
\begin{tabular}{lcccccc}
& \multicolumn{3}{c}{\textbf{$R^2$}} & \multicolumn{3}{c}{\textbf{$R^2_{adj}$}} \\
\cmidrule(r){2-4} \cmidrule(l){5-7} 
\textbf{Target } & $\mathbf{\gamma_{ij}=0}$ & $\mathbf{\gamma_{ij} \neq 0}$ & \(\delta\) & $\mathbf{\gamma_{ij}=0}$ & $\mathbf{\gamma_{ij} \neq 0}$ & \(\delta\) \\

\midrule
proj4 & 0.758 & \textbf{0.855} & 0.097 & 0.757 & \textbf{0.847} & 0.090 \\
json & 0.629 & \textbf{0.853} & 0.224 & 0.628 & \textbf{0.848} & 0.220 \\
curl & 0.672 & \textbf{0.820} & 0.148 & 0.672 & \textbf{0.814} & 0.142 \\
re2 & 0.586 & \textbf{0.817} & 0.231 & 0.585 & \textbf{0.811} & 0.226 \\
freetype & 0.639 & \textbf{0.684} & 0.045 & 0.638 & \textbf{0.658} & 0.020 \\
bloaty & 0.524 & \textbf{0.616} & 0.092 & 0.522 & \textbf{0.593} & 0.071 \\
php & 0.576 & \textbf{0.613} & 0.037 & 0.575 & \textbf{0.595} & 0.020 \\
libxml & 0.472 & \textbf{0.519} & 0.047 & 0.478 & \textbf{0.494} & 0.016 \\
sqlite & 0.350 & \textbf{0.380} & 0.030 & 0.349 & \textbf{0.350} & 0.001 \\
\bottomrule
\end{tabular}
\label{tab:results_rq1}
\end{table}

The intuition is that the mutator $i$ in the sequence $\langle i,j \rangle$ is associated with a constant effect $\alpha_i$ (or $\beta_j$ for mutator $j$).
The effect $a_i$ means that, whenever $i$ is the first mutator in the sequence, we should expect an increase (or decrease) in the number of interesting inputs (relative to the bias $\mu$) of $\alpha_i$ inputs.
The same holds for $\beta_j$. 
Regarding $\gamma_{ij}$, it denotes the interaction effect of mutators $i$ and $j$ and is interpreted as follows:
Suppose that we expect the mutator $i$ in the first position of the sequence (independent of the second) to increase the number of interesting inputs by $\alpha_i$ and the mutator $j$ in the second position of the sequence (independent of the first) to increase the number of interesting inputs by $\beta_j$.
Now, suppose that when $i$ is combined with $j$ we observe a significant increase (or decrease) by $\gamma_{ij} \neq 0$; this would mean that there exists an interaction effect between $i$ and $j$.
If, on the contrary, $\gamma_{ij}=0$, it would mean that there is no interaction effect.

To test the significance of the interaction term $\gamma_{ij}$, we use the two-way ANOVA~\cite{anova}, which tests the null hypothesis that $\gamma_{ij}=0$ for all $i,j \in \mathcal{M}$. 
ANOVA has the following three assumptions~\cite{anova_assumptions}:
First, that the dataset observations are independent.
Since AFL++ saves interesting inputs to be further mutated in the future, possibly from a different pair of mutators, the independence assumption could be violated.
To test the independence assumption, we employ the \textit{Residuals vs Fitted} plot~\cite{linear_assumptions}.
In nine target programs, the residuals follow random patterns, indicating that the independence assumption is met~\cite{anova_assumptions}, while in four target programs the residuals exhibit a systematic trend, indicating that the independence assumption is violated.
We do not fit a model in these four programs.
Second, ANOVA assumes that the residuals are normally distributed, which we check by plotting the residuals and observing a bell-shaped distribution in all nine programs~\cite{anova_assumptions}.
Finally, ANOVA assumes homoscedasticity, which we also check using the \textit{residuals-vs-fitted} plot~\cite{anova_assumptions}.
We find heteroscedasticity in 3/9 programs, so we use the robust \texttt{hc3} error~\cite{robust_hc3} to mitigate it. 

Complementary to the ANOVA results, we calculate the goodness-of-fit, measured with $R^2$ and $R^2_{adj}$~\cite{DraperSmith1998}, of the linear model with the interaction term ($\gamma_{ij} \neq 0$) and a linear model without the interaction term ($\gamma_{ij}=0$).
A higher goodness-of-fit for the variation with the interaction term ($\gamma_{ij} \neq 0$) would indicate the interaction term affects the number of interesting inputs.

Our study focuses on the interactions between mutator types, but it is possible that interactions also exist with other factors, such as mutator locations (i.e., which part of the input is mutated).
There is no reason to believe that the mutation locations or other factors are not uniformly distributed across mutator pairs. 
Combined with the fact that the number of mutations in our experiments is in the order of millions, we can reasonably expect that the effect of extraneous variables such as location evens out across all mutator pairs.
More formally, in our experiment, the independent variables (i.e., the factors being manipulated) are the mutator types, the dependent variable (i.e., the factor being measured) is the number of interesting inputs, and extraneous variables are other factors that could potentially influence the dependent variable but are \textit{not manipulated}, and as such, they pose no threat for the validity of the observed interaction effect.

\subsection{Results}\label{ssec:motiv_results}
To test the null hypothesis that $\gamma_{ij}=0$ for all $i,j \in \mathcal{M}$, we run a two-way ANOVA on the fitted model (one model for each target program).
For all target programs we get $p < 0.0001$, hence reject the null hypothesis.
This should not be misinterpreted as \emph{all} mutator pairs having a significant interaction effect, rather that there exist \emph{some} mutator pairs with a significant interaction effect.
Although two-way ANOVA also generates p-values for each of the individual hypotheses $\gamma_{ij}=0$ for fixed $i$ and $j$, using these p-values to draw inferences is a bad practice because $|\mathcal{M}|^2=1024$ hypotheses are tested simultaneously, hence suffering from \emph{multiple hypothesis testing}~\cite{multiple_hypothesis_testing}.
In \Cref{sec:discussion} we manually investigate which mutators have strong interaction effects.

Complementary to the ANOVA analysis, we show the goodness-of-fit in \Cref{tab:results_rq1}. 
We focus on $R^2_{adj}$ because it penalizes the number of model parameters, which is higher when $\gamma_{ij} \neq 0$, but even with this penalization we see that the model with the interaction term achieves higher $R^2_{adj}$.
This means that the interaction term affects the number of interesting inputs, which agrees with the ANOVA null hypothesis test.

\vspace{10pt}
{
\setlength{\fboxsep}{7pt}

\noindent
\fcolorbox{black}{gray!8}{
  \parbox{0.44\textwidth}{
    \textbf{Finding 1}. 
    We measure an interaction effect between two mutators on the number of interesting inputs.
  }
}
}
\vspace{10pt}

\section{\ourtool}\label{sec:methodology}

The establishment of the existence of an interaction effect between two mutators does not guarantee the practical significance of this effect. 
To investigate its value for fuzzing, \textbf{we propose a method for generating mutator sequences where the probability of selecting the next mutator is conditioned on the previously selected one}.
We implement this method into \ourtool and compare its performance against state-of-the-art fuzzers.

\subsection{Definitions and Problem Statement}
The goal of a mutation strategy is to mutate a seed $s$ by applying a sequence of mutators.
This sequence is of the form $\langle m_1, m_2, \cdots, m_l \rangle$.
The mutators $m_n$ are selected from a predefined set of mutators $\mathcal{M} = \{1, 2, ..., |\mathcal{M}|\}$.
For example, AFL++ defines $|\mathcal{M}|=32$ mutators which are shown in \Cref{tab:mutators}.

We reduce the problem of generating a mutator sequence \linebreak
$\langle m_1, m_2, \cdots, m_l \rangle$ 
to the problem of learning the conditional probability $\Pr(m_{n}=j \mid m_{n-1}=i)$ of selecting $j$ as the next mutator in the sequence given that the previously selected mutator is $i$.
If we learn such a probability, we can iteratively generate a mutator sequence of arbitrary length $l$ with the following steps:
We randomly select the first mutator $m_1 \in \mathcal{M}$.
Then, \emph{for fixed} $m_1$, we obtain $m_2$ by sampling from the distribution
\[m_2 \sim \Pr(j) = \Pr(m_2 =j \mid m_1) \] 
In general, we obtain
$m_n$ by sampling from the distribution 
\[
m_n \sim \Pr(j) = \Pr(m_n = j \mid m_{n-1})
\]
where $m_{n-1}$ is fixed from the previous step.
This sampling algorithm is also known as a Markov random walk~\cite{markov} on the Markov chain, where the states are the mutators and $\Pr(m_{n}=j \mid m_{n-1}=i)$ are the transition probabilities from state $i$ to state $j$.

\begin{figure*}
  \centering
  \includegraphics[width=0.99\textwidth]{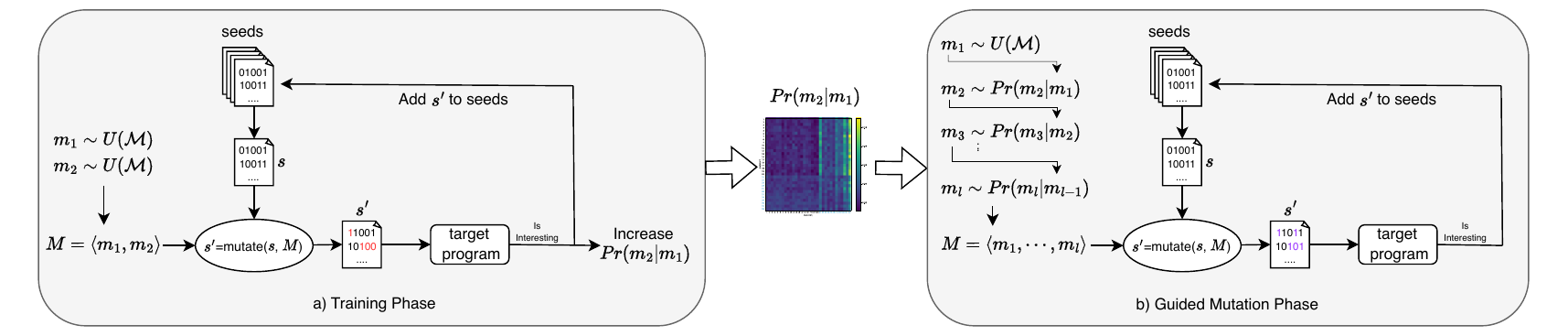}
  \caption{High-level overview of \ourtool.}
  \Description{High-level overview of our proposed fuzzer.}
  \label{fig:method}
\end{figure*}

\subsection{Overview}
Our mutation strategy consists of two phases.
During the \textbf{training phase} (\Cref{fig:method}a), we aim to learn the conditional probability $\Pr(m_{n}=j \mid m_{n-1}=i)$ of selecting $j$ as the next mutator in the sequence given that the previously selected mutator is $i$.
We model this probability by computing the number of interesting inputs produced by all possible mutator pairs $\langle i,j \rangle$ in the first $T_{train}$ hours of fuzzing.
Then, for fixed $i$, we define $\Pr(m_{n}=j \mid m_{n-1}=i)$ to be proportional to the number of interesting inputs produced by $\langle i,j \rangle$.

After the first $T_{train}$ hours, our fuzzer enters the \textbf{guided mutation phase} (\Cref{fig:method}b).
Here, it generates mutator sequences by sampling from the learned distribution $\Pr(j)=\Pr(m_{n}=j \mid m_{n-1}=i)$ where the mutator selected in each step is used to select the mutator of the next step.
The training time $T_{train}$ that \ourtool spends in the training phase before switching to the guided mutation phase is a hyperparameter that we finetune in \Cref{sec:evaluation}.

\subsection{Training Phase}\label{ssec:methodology_training}
The goal of the training phase is to learn the conditional probability $\Pr(m_{n}=j \mid m_{n-1}=i)$ of selecting $j$ as the next mutator in the sequence given that the previously selected mutator is $i$.
\emph{We define this probability to be proportional to the number of interesting inputs generated by the sequence $\langle i,j \rangle$}; 
we denote this number with $N(i,j)$.
This design decision is intuitive: 
we give a higher probability of selecting the mutator $j$ if $j$ has generated more interesting inputs when applied on top of mutator $i$.
To obtain $N(i,j)$, in the first $T_{train}$ hours we run AFL++ with the three modifications \textbf{M1}, \textbf{M2}, and \textbf{M3} we applied in \Cref{sec:rq1}:

\begin{description}
    \item[\textbf{M1}:] We limit the length of the mutator sequences to $l=2$.

    \item[\textbf{M2}:] We sample the mutators uniformly at random.
    
    \item[\textbf{M3}:] We update $N(i, j)$.
\end{description}

The modifications \textbf{M1}–\textbf{M3} are the only changes we apply to \aflpp, while all other parameters remain at their default values.
After $T_{train}$ hours, we execute the last step of the training phase: we define the probability of selecting $j$ as the next mutator in the sequence given that the previously selected mutator is $i$
as \begin{equation}\label{eq:1}
\Pr(j) = \Pr(m_{n}=j \mid m_{n-1}=i) = \frac{N(i, j)}{\sum_{j=1}^{|\mathcal{M}|} N(i, j)}
\end{equation}

The division with the sum turns each row into a probability distribution that sums to $1$.
The distribution for the \texttt{freetype} program is shown on the right side of \Cref{fig:matrices}, while the left side shows the associated $N(i, j)$.
This concludes the training phase and \ourtool then automatically enters the guided mutation phase.
We note that our training phase is lightweight and embedded in the fuzzing loop of AFL++.
This is in contrast to machine learning approaches where AFL++ is run for one hour to gather data, which are then used to train a neural network~\cite{neuzz,mtfuzz}.
Our training is as simple as increasing $N(i, j)$ whenever an interesting input is generated.

\subsection{Guided Mutation Phase}\label{ssec:methodology_guided_mutation}

\begin{algorithm}
\LinesNumbered
\scriptsize
\caption{Guided Mutation Phase}
\label{alg:guided_mutation_phase}
\KwIn{subroutines $select\_sequence\_length()$ and $select\_first\_mutator()$}
\KwOut{mutator sequence $\langle m_1, m_2, ..., m_l \rangle$}
$l \gets select\_sequence\_length()$\;
$m_1 \gets select\_first\_mutator()$\;
$M \gets \emptyset$\;

\For{$n = 2$ \KwTo $l$}{
    $m_n \sim \Pr(m_n \mid m_{n-1})$\;
    $M \gets M \cup \{m_n\}$\;
}

\Return{$M = \langle m_1, m_2, ..., m_l \rangle$}\;

\end{algorithm}

The goal of the guided mutation phase is to generate mutator sequences $\langle m_1, m_2, ..., m_l \rangle$ by leveraging the learned probability distribution of \Cref{eq:1}.
We propose \Cref{alg:guided_mutation_phase} to generate a sequence $\langle m_1, m_2, ..., m_l \rangle$ and replace the \colorbox{myblueopaque}{highlighted} part of \Cref{alg:greybox_fuzzing}:
First, we select the sequence length $l$ (line 1) and the first mutator $m_1$ (line 2).
The subroutines that perform these selections are hyperparameters and are discussed in detail below.
Then, at each step $2<n<l$ (line 4), we obtain $m_n$ by sampling from the learned probability $\Pr(m_n \mid m_{n-1})$ (line 5), where now $m_{n-1}$ is fixed from the previous step.
In the remainder of this section, we discuss the hyperparameters of the guided mutation phase, and in \Cref{sec:evaluation} we experimentally select the best-performing values.
By default, \ourtool uses the best-performing hyperparameters as per \Cref{sec:evaluation} unless stated otherwise.

\subsubsection{Selecting the Sequence Length}
\Cref{alg:guided_mutation_phase} leaves open the question of what the optimal length $l$ of a mutator sequence is.
We consider two options for this hyperparameter.
The \textit{first option} is to keep the AFL++ default sequence length that ranges from \num{2} to \num{16}.
However, due to our optimized generation algorithm, a different, potentially smaller length may be sufficient.
For this reason, we propose a \textit{second option}: treating the sequence length as a dynamically determined variable. 
We follow previous work~\cite{slopt} and model the sequence length selection as a multi-armed bandit (MAB) problem~\cite{mab}.
We solve this problem using the $\varepsilon$-greedy algorithm, a well-known heuristic solution to the MAB problem that has been applied in previous work~\cite{egreedy}. 
Given a set of possible values for the sequence length $l \in L=\{2, 3, ..., 16\}$, the $\varepsilon$-greedy algorithm selects the best performing (i.e., the one that has generated the most interesting inputs) sequence length with probability $\varepsilon$, and with probability $1-\varepsilon$ it selects the sequence length uniformly at random from $L$.
The value of $\varepsilon$ follows a step decay, starting from $\varepsilon=1$ (exploration only) in the first hour of the guided mutation phase, then dropping to $\varepsilon=0.5$ to boost exploitation.
We select the $\varepsilon$-greedy algorithm over other heuristic MAB algorithms like Upper Confidence Bound (UCB)~\cite{ucb} or Thompson Sampling~\cite{mab} because of its simplicity and interpretability.

\subsubsection{Selecting the First Mutator}
Another hyperparameter of the guided mutation phase is how to select the first mutator $m_1$.
We investigate two options for $m_1$:
\begin{enumerate}[label=(\alph*)]
    \item a uniformly random selection of $m_1$ from $\mathcal{M}$, as is usually done in random walks~\cite{markov}.
    \item a weighted selection of $m_1$ that gives a higher probability to mutators that produced more interesting inputs.
\end{enumerate}

To implement (b), for each mutator $i$ we calculate a score 
\[
s_i = \sum_{j=1}^{|\mathcal{M}|}N(i, j) + \sum_{j=1}^{|\mathcal{M}|}N(j, i) - N(i, i)
\]
that holds the total number of interesting inputs produced by mutator $i$ during the training phase.
The first sum corresponds to the number of interesting inputs where $i$ was the first mutator in the sequence, the second sum corresponds to the number of interesting inputs where $i$ was the second mutator in the sequence, and we subtract the last term because $N(i, i)$ was counted twice, one time in each sum.
After computing $s_i$  for each $i \in \mathcal{M}$, we normalize them by dividing by their sum to form a probability distribution and sample the first mutator $m_1$ from that distribution.

\subsubsection{Using More Mutators as Context}\label{sssec:more_mutators}
The decision to use only the last mutator instead of the last $p>1$ mutators is not arbitrary.
Modeling the probability $Pr\left( m_n \mid m_{n-1}m_{n-2}...m_{n-p}  \right)$ of selecting the mutator $m_n$ given the previous $p$ mutators is possible, but it suffers from the curse of dimensionality~\cite{koppen2000curse}:
The 2-dimensional matrix $N(i, j)$ would become $p+1$-dimensional.
Even for $p=2$, we would end up with $32^3=32768$ unique mutator triplets.
Due to this combinatorial explosion, most triplets remain unobserved during training because they do not produce a coverage-increasing input.
This guided our design to look only at the previous mutator when selecting the next one.
Nevertheless, we run experiments with $p=2$ in \Cref{sssec:evaluation_using_more_muts} to validate our reasoning.

\subsection{Alternative Designs}
In our two-phase design, the probability learned in the first $T_{train}$ hours of fuzzing guides the mutations for the remainder of the fuzzing campaign, which, as we show below, is effective in most cases.
However, alternative designs that move the training phase later could be considered, and we discuss two of them here, explaining why they were not preferred.
The first alternative would be to dynamically update $N(i, j)$ throughout the entire fuzzing campaign in a Multi-Armed Bandit (MAB) setting~\cite{mab}, as done in similar fuzzing optimizations~\cite{havocMAB,egreedy}.
However, we showed in RQ1 that there exist interactions between mutators, which the MAB setting cannot handle because it assumes independence of mutators (arms).
Another alternative would be to move or repeat the training phase later in the fuzzing campaign, because some mutator pairs could be more effective early on but less so toward the end.
However, this design suffers from the limited amount of training data (i.e., interesting inputs discovered) since greybox fuzzers discover new interesting inputs near-logarithmically~\cite{exponential_cost}. We experimentally validated this in initial experiments by shifting the training phase one hour later in the fuzzing campaign and observing a significant performance drop.

\section{Empirical Evaluation Of \ourtool}\label{sec:evaluation}
We develop our novel mutation strategy into a fuzzer named \ourtool by replacing the mutation strategy of AFL++ and compare its performance to state-of-the-art baselines.
Since achieved code coverage is the most widely used measure of fuzzer performance~\cite{sok}, we compare the code coverage of \ourtool against the baselines to answer our first research question.

\vspace{10pt}
{
\setlength{\fboxsep}{7pt}
\noindent
\fcolorbox{black}{gray!8}{
  \parbox{0.44\textwidth}{
    \textbf{RQ2}. 
    How does \ourtool compare to AFL++ and MOPT in terms of code coverage?
  }
}
}
\vspace{10pt}

Coverage alone is not a sufficient measure of performance~\cite{inozemtseva2014coverage,sok,klees2018evaluating}.
It is only useful when accompanied by other measures, mainly the number of bugs found. Our next research question compares the number of bugs found by \ourtool against the baselines.

\vspace{10pt}
{
\setlength{\fboxsep}{7pt}
\noindent
\fcolorbox{black}{gray!8}{
  \parbox{0.44\textwidth}{
    \textbf{RQ3}. 
    How does \ourtool compare to AFL++ and MOPT in terms of bugs found?
  }
}
}
\vspace{10pt}

Finally, we follow recent work on fuzzer evaluation~\cite{sok,klees2018evaluating,revisitingNPS} that strongly recommends performing an ablation study when evaluating a new fuzzer for two main reasons.
First, to increase the confidence that the success of \ourtool comes from leveraging the information about the previously selected mutator and is not a byproduct of some latent design decision.
Second, to understand how different hyperparameters affect the performance.
For these reasons, we design an ablation study to extensively evaluate \ourtool under different hyperparameters.

\vspace{10pt}
{
\setlength{\fboxsep}{7pt}
\noindent
\fcolorbox{black}{gray!8}{
  \parbox{0.44\textwidth}{
    \textbf{RQ4}. 
    How do different hyperparameters affect the performance of \ourtool?
  }
}
}
\vspace{10pt}

\subsection{Experimental Setup}
The empirical comparison of two fuzzers is a challenging problem, mainly due to the inherent randomness of fuzzing.
Klees et al.~\cite{klees2018evaluating} propose that the fuzzers run for \num{24} hours and the experiment is repeated \num{20} to \num{30} times for a single program.
Then, statistical tests like the Mann-Whitney U test~\cite{MannWhitney} should be performed to test the hypothesis that one fuzzer achieves higher coverage than the other, and the p-values should be reported along with the effect size.

\subsubsection{Implementation}
We implement \ourtool on top of the latest version of AFL++ at the time, which was \num{4.21}a.
Hence, the baselines are also versioned on AFL++ \num{4.21}a.
We open source our implementation to be independently assessed by the community~\cite{ours}.

\subsubsection{Baselines}
A best practice when selecting baselines is to always compare against the fuzzer on top of which the new fuzzer is built~\cite{sok,klees2018evaluating}.
In our case, we build on top of AFL++, so this is our first baseline.
Additionally, it is suggested to compare against state-of-the-art fuzzers that propose different approaches to the same problem, in our case the mutation strategy.
MOPT~\cite{mopt} proposes a mutation strategy that optimizes the selection probability of each mutator \emph{in isolation}, in contrast to our approach which optimizes the probability of mutator pairs, which makes it a well-suited baseline.
Moreover, Schloegel et al.~\cite{sok} found that MOPT is consistently selected as a baseline lately, so we select MOPT as a second baseline.
We use the AFL++ version of MOPT, since the original implementation of MOPT is based on AFL, which would put MOPT at a disadvantage.
We intended to use SeamFuzz~\cite{seamfuzz} as an additional baseline for the same reason as MOPT, but their replication package is based on an old version of AFL++ v3.15, which would also put SeamFuzz at a great disadvantage.
After our inquiry to the authors, they assured us that they are working on porting SeamFuzz to the latest version in the near future.

\subsubsection{Benchmarks}\label{sssec:benchmarks}
The most widely used fuzzing benchmark is FuzzBench~\cite{fuzzbench,sok}, which contains a diverse set of real-world open-source C programs representing a variety of application domains. 
FuzzBench automates the execution and coverage measurement, ensuring reproducible and consistent results.
For these reasons, we use FuzzBench in our experiments.
However, FuzzBench does not report the number of bugs found.
Hence, we also use MAGMA~\cite{magma}, a benchmark of real-world bugs that MAGMA developers manually ported to the latest version of each program.

\subsubsection{Infrastructure}
All experiments take place in \num{16} identical virtual machines (VMs) running Ubuntu \num{22.04}, each having an AMD EPYC 7702 processor with \num{32} CPUs, \num{125} GB RAM, and a \num{100} GB disk.
We restrict experiments of the same target program to the same set of VMs, for example, we use only VM1 and VM2 for the experiments on \texttt{proj4} to further ensure fairness (although the VMs are identical).
Only \num{24} out of \num{32} cores (\(\sim\)\num{80}\%) are used simultaneously.

\begin{table}
\scriptsize
\centering
\caption{Median coverage and st.d. after 24 hours of \ourtool (ours), \aflpp, and \mopt on FuzzBench. 
}
\begin{tabular}{l@{\hskip 0.05in}|@{\hskip 0.05in}r@{\hskip 0.03in}c@{\hskip 0.03in}r@{\hskip 0.05in}|@{\hskip 0.05in}r@{\hskip 0.03in}c@{\hskip 0.03in}r@{\hskip 0.05in}|@{\hskip 0.05in}r@{\hskip 0.03in}c@{\hskip 0.03in}r@{\hskip 0.05in}|@{\hskip 0.05in}l@{\hskip 0.03in}c@{\hskip 0.05in}|@{\hskip 0.05in}l@{\hskip 0.03in}c@{\hskip 0.05in}}
\textbf{Target} & \multicolumn{3}{c}{\textbf{AFL++}} & \multicolumn{3}{c}{\textbf{MOPT}} & \multicolumn{3}{c}{\textbf{\ourtool}} & $\mathbf{p_{AFL++}}$ & $\mathbf{\hat{A}_{12}}$ & $\mathbf{p_{MOPT}}$ & $\mathbf{\hat{A}_{12}}$ \\
\midrule
bloaty   & \underline{5,990} & $\pm$  & 78 & 5,216 & $\pm$ & 332 & \textbf{6,028} & $\pm$ & 73 & 0.0185** & 0.65 & 0.0000*** & 0.99 \\
curl     & \underline{10,867} & $\pm$ & 102 & 10,773 & $\pm$ & 94 & \textbf{10,912} & $\pm$ & 82 & 0.0126** & 0.67 & 0.0000*** & 0.89 \\
freetype & 11,006 & $\pm$ & 484 & \textbf{11,682} & $\pm$ & 327 & \underline{11,491} & $\pm$ & 397 & 0.0006*** & 0.73 & 0.9987 & 0.31 \\
json     & \textbf{520} & $\pm$ & 0 & \textbf{520} & $\pm$ & 1 & \textbf{520} & $\pm$ & 0 & NaN & NaN & 0.0926* & 0.56 \\
lcms     & 1,956 & $\pm$  & 204 & \underline{2,002} & $\pm$ & 218 & \textbf{2,004} & $\pm$ & 348 & 0.3482 & 0.52 & 0.9525 & 0.40 \\
libpcap  & \underline{2,802} & $\pm$  & 125 & 2,780 & $\pm$ & 113 & \textbf{2,840} & $\pm$ & 131 & 0.0323** & 0.61 & 0.0077** & 0.64 \\
libpng   & 2,004 & $\pm$  & 3 & \underline{2,005} & $\pm$ & 26 & \textbf{2,006} & $\pm$ & 3 & 0.0781* & 0.59 & 0.2547 & 0.53 \\
libxml   & 19,212 & $\pm$ & 244 & \underline{19,247} & $\pm$ & 891 & \textbf{19,256} & $\pm$ & 215 & 0.7313 & 0.46 & 0.2214 & 0.55 \\
openssl  & 5,818 & $\pm$  & 7 & \textbf{5,828} & $\pm$ & 6 & \underline{5,827} & $\pm$ & 5 & 0.0000*** & 0.82 & 0.8350 & 0.42 \\
php      & \textbf{16,712} & $\pm$ & 79 & 16,653 & $\pm$ & 23 & \underline{16,689} & $\pm$ & 90 & 0.4923 & 0.50 & 0.0147** & 0.69 \\
proj4    & 6,872 & $\pm$  & 134 & \underline{6,981} & $\pm$ & 179 & \textbf{7,068} & $\pm$ & 175 & 0.0000*** & 0.78 & 0.0559* & 0.59 \\
re2      & \underline{2,876} & $\pm$  & 4 & 2,874 & $\pm$ & 5 & \textbf{2,879} & $\pm$ & 5 & 0.0022*** & 0.68 & 0.0002*** & 0.75 \\
sqlite   & \underline{19,907} & $\pm$ & 801 & 18,954 & $\pm$ & 798 & \textbf{20,091} & $\pm$ & 254 & 0.0003*** & 0.76 & 0.0000*** & 0.97 \\

\bottomrule
\multicolumn{14}{r}{Highest median = \textbf{bold}, second highest = \underline{underlined}. ***: $p<0.01$, **: $p<0.05$, *: $p<0.1$.} \\

\end{tabular}
\label{tab:results_rq2}
\end{table}

\begin{table}
    \centering
    \scriptsize
    \caption{Time needed (hours) for \ourtool to reach the coverage achieved by the baselines after 24 hours.}
    \begin{tabular}{lrrrrrr}
        & \multicolumn{3}{c}{\textbf{Time to final AFL++ cov.}} & \multicolumn{3}{c}{\textbf{Time to final MOPT cov.}} \\
        \cmidrule(lr){2-4} \cmidrule(lr){5-7}
        \textbf{Target} & By \aflpp & By \ourtool & $\delta$ & By \mopt & \ourtool & $\delta$ \\
        \midrule
        bloaty    & 23.8 & 20.0   & \textbf{3.8}  & 23.5 & 1.5  & \textbf{22.0}   \\
        curl      & 22.5 & 15.8 & \textbf{6.7}  & 23.8 & 7.0    & \textbf{16.8} \\
        freetype  & 23.8 & 12.0   & \textbf{11.8} & 17.8 & 23.8 & -6.0   \\
        lcms      & 23.5 & 13.8 & \textbf{9.7}  & 23.0   & 21.5 & \textbf{1.5}  \\
        libpcap   & 23.8 & 19.2 & \textbf{4.6}  & 23.8 & 16.5 & \textbf{7.3}  \\
        libpng    & 17.2 & 7.8  & \textbf{9.4}  & 19.2 & 11.0   & \textbf{8.2}  \\
        libxml    & 23.8 & 22.8 & \textbf{1.0}    & 23.8 & 23.2 & \textbf{0.6}  \\
        openssl   & 15.5 & 9.0    & \textbf{6.5}  & 18.2 & 19.5 & -1.3 \\
        php       & 22.0   & 23.8 & -1.8 & 23.8 & 19.8 & \textbf{4.0}    \\
        proj4     & 23.8 & 14.2 & \textbf{9.6}  & 23.8 & 18.5 & \textbf{5.3}  \\
        re2       & 23.8 & 13.5 & \textbf{10.3} & 21.5 & 9.2  & \textbf{12.3} \\
        sqlite    & 23.8 & 19.0   & \textbf{4.8}  & 22.0   & 8.8  & \textbf{13.2} \\

        \midrule
        Median    & 23.8 & 15.0   & 6.6           & 23.3 & 16.5 & 6.3 \\
        \bottomrule
    \end{tabular}
    \label{tab:latency}
\end{table}

\subsection{RQ2 --- Code Coverage}\label{ssec:evaluation_rq2}

FuzzBench contains \num{28}  programs but is continuously maintained so some  programs that existed in older versions may not exist anymore.
Evaluating in all \num{28} programs is computationally intensive: as reported by Schloegel et al.~\cite{sok}, fuzzers in top venues of the last five years are evaluated on \num{8.9} programs on average.
To avoid bias in the program selection, we use the same programs as two recent fuzzing papers that also used FuzzBench~\cite{seamfuzz,dipri}.
Five of the programs used by Li et al.~\cite{seamfuzz} are available in the latest FuzzBench version, as well as eight of the programs used by Qian et al.~\cite{dipri}.
We use their union, totaling thirteen programs.

We run each fuzzer on each program for \num{24} hours and \num{30} trials while keeping the default initial seeds of FuzzBench.
We report the median coverage and the standard deviation after \num{24} hours in \Cref{tab:results_rq2}.
To compare \ourtool with a baseline, we employ the Mann-Whitney U test with the null hypothesis that \ourtool does not achieve higher coverage than the baseline.
Since we have two baselines, AFL++ and MOPT, we run the test two times, one against AFL++ and one against MOPT.
The resulting p-values, $p_{AFL++}$ and $p_{MOPT}$ respectively, are also shown in \Cref{tab:results_rq2}, along with the associated Vargha-Delaney effect size~\cite{vargha_delaney} ($\hat{A}_{12}$), which denotes the probability that a random trial of \ourtool will achieve higher coverage than a random trial of the baseline fuzzer.
We see that \textsc{MuoFuzz} achieves higher coverage than AFL++ in 9/13 programs and than MOPT in 8/13 programs with statistical significance.

\Cref{tab:results_rq2} answers the question of how many more branches can \ourtool discover in a 24-hour campaign, but does not tell us anything about how faster it reaches that coverage compared to the baselines.
This is particularly important as Böhme et al.\cite{exponential_cost} found that the cost of discovering new branches in greybox fuzzers increases \textit{exponentially} with time.
In other words, even a seemingly small increase in the final coverage after 24 hours may be hard to achieve.
Hence, they suggest a complementary metric when comparing coverage between two fuzzers: How long does it take one fuzzer to reach the final coverage the other fuzzer achieves after 24 hours?
For example, after \num{24} hours in the \textit{curl} program, \aflpp achieves a median coverage of \num{10867} while \ourtool achieves \num{10912}, which may seem like a not-so-important increase at first glance (\num{0.9}\%). However, \ourtool reaches \num{10867} coverage after only \num{15.8} hours (median over \num{30} trials), while \aflpp first reaches that number after \num{22.5} hours (its median coverage does not increase in the last \num{1.5} hours).
We provide this metric for all target programs in \Cref{tab:latency}.
We find that \ourtool reaches the 24-hour-coverage of \aflpp after \num{15} hours (median across all programs) and the 24-hour-coverage of \mopt after \num{16.5} hours.

To better understand when \textsc{MuoFuzz} yields coverage improvements, we analyzed the programs in which \textsc{MuoFuzz} did not achieve a statistically significant improvement in coverage over either of the baselines.
For the \texttt{lcms}, \texttt{libpng}, and \texttt{openssl} programs, the interaction effect could not be observed already from our empirical analysis in RQ1: 
The data for these three programs violated the independence assumption and thus we did not run the ANOVA analysis. 
Moreover, for \texttt{libpng} and \texttt{openssl}, the learned probability matrix does not show clear underlying patterns, in contrast to the other programs where such patterns are visible, indicating the absence of an interaction effect.
For \texttt{freetype}, \texttt{libxml}, and \texttt{php}, the goodness-of-fit ($R^2_{adj}$) in Table 2 is 0.658, 0.595, and 0.494 respectively, indicating that the linear model of Equation (1) only partially captures $N(i, j)$ and, as a consequence, the interaction effect may not be strong. 
In all programs where $R^2_{adj}>0.8$, \textsc{MuoFuzz} outperforms both baselines with statistical significance, except for AFL++ in \texttt{json}.
For \texttt{json}, all fuzzers reach the maximum coverage of 520 after only 15 minutes, except for some runs of MOPT. This early saturation suggests that the mutation strategy does not affect the coverage in the \texttt{json} program and that other techniques, such as symbolic execution, may be required to explore \texttt{json} in more depth.

\vspace{10pt}
{
\setlength{\fboxsep}{7pt}
\noindent
\fcolorbox{black}{gray!8}{
  \parbox{0.44\textwidth}{
    \textbf{Finding 2}. \textsc{MuoFuzz} achieves higher coverage than AFL++ and MOPT in 9/13 and 8/13 programs respectively, while for the remaining programs we observed a weak interaction effect that did not lead to improved performance. \textsc{MuoFuzz} needs 15 and 16.5 hours on average to reach the 24-hour coverage of AFL++ and MOPT respectively.
  }
}
}
\vspace{10pt}

\subsection{RQ3 --- Bugs Found}

In the second set of experiments we compare the bug-finding ability of \ourtool to the baseline fuzzers.
The latest version of MAGMA (v.1.2) comes with eight target programs, so we use all of them.
A target program may have more than one driver file in MAGMA, in which case we fuzz all of them.
The list of programs along with their driver files is available in the MAGMA \href{https://github.com/HexHive/magma}{repository}.

We run each fuzzer on each benchmark for \num{24} hours and \num{20} trials, while keeping the default initial seeds of MAGMA.
\ourtool detects four bugs that AFL++ cannot detect (\texttt{XML002}, \texttt{SSL009}, \texttt{LUA002}, and \texttt{SQL010}) while AFL++ detects only one bug that \ourtool cannot detect (\texttt{TIF001}).
In other words, \ourtool finds three more bugs than AFL++.
When we look at \texttt{TIF001}, we see that neither \textsc{MuoFuzz} nor MOPT found the bug. Only 1 of the 20 trials of the baseline AFL++ found it, quite possibly due to the inherent randomness of fuzzing. In contrast, \textsc{MuoFuzz} found the three bugs that AFL++ did not find in 3/20, 3/20, and 2/20 trials respectively.
Regarding MOPT, it finds the same number of bugs as \ourtool, but \texttt{SSL001} is only found by MOPT and \texttt{SQL010} is only found by \ourtool.
This suggests that although the total number of bugs is the same, \ourtool may trigger a different, unique behaviour compared to MOPT.
The bug \texttt{SSL001} that \textsc{MuoFuzz} missed belongs to the \textit{openssl} program, where the interaction effect is weak and no clear pattern is evident, as shown in~\Cref{ssec:evaluation_rq2}.

We also compare the number of unique bugs triggered by each fuzzer averaged across the \num{20} trials.
In the \texttt{libsndfile} program, all three fuzzers trigger exactly seven bugs in all trials. In the remaining seven programs, \ourtool ranks first in four, \mopt ranks first in two, and \aflpp ranks first in one.
However, the differences in the average number of unique bugs are not statistically significant according to the Mann-Whitney U test~\cite{MannWhitney}. 
We see a similar pattern in the time-to-bug (considering the bugs that all three fuzzers found), where no single fuzzer consistently outperforms the other two.
The time-to-bug as well as the number of average unique bugs for every fuzzer and program are provided in our replication package for space reasons~\cite{ours}.

\vspace{10pt}
{
\setlength{\fboxsep}{7pt}
\noindent
\fcolorbox{black}{gray!8}{
  \parbox{0.44\textwidth}{
    \textbf{Finding 3}. \textsc{MuoFuzz} detects three more bugs than AFL++. It also detects the same number of bugs as MOPT, but each fuzzer discovered a distinct bug, with \textsc{MuoFuzz} missing a bug in a program where the interaction effect was weak.
    The above suggest that leveraging the interaction effect, when it exists, can trigger unique program behaviour.
}
}
}
\vspace{10pt}

To better understand how the design choices and hyperparameters of \textsc{MuoFuzz} affect its performance across different target programs, we conduct a detailed ablation study in~\Cref{ssec:rq3_methodology}.

\subsection{RQ4 --- Ablation Study}\label{ssec:rq3_methodology}

The goal of the ablation study is to better understand the effect of various components and hyperparameters of the proposed fuzzer.
Since fuzzers are complex systems comprised of many components, some components may contribute most to the performance while others do not contribute at all.
In the rest of this section, we run experiments to understand how different hyperparameters of \ourtool contribute to its performance.
We focus on FuzzBench because it yielded more decisive results in the first set of experiments.
We report the results in \Cref{tab:results_ablation}.

\subsubsection{Using More Mutators as Context}\label{sssec:evaluation_using_more_muts}
\ourtool takes into account only the last mutator when predicting the next.
We reasoned about the disadvantages of using $p>1$ previous mutators in \Cref{sssec:more_mutators};
we experimentally validate this reasoning by implementing a variation of \ourtool that takes into account $p=2$ previous mutators to select the next one.
We modify the logic as described in \Cref{sssec:more_mutators} and call this variation \ourtool{}$_{p=2}$.
\Cref{tab:results_ablation} shows that \ourtool{}$_{p=2}$ underperforms \ourtool in all target programs.
By manually analyzing $N(i, j, k)$ for each program, we see that the sparsity~\cite{sparse} ranges from $0.003$ to $0.1$, which validates that the learning suffers from the curse of dimensionality~\cite{koppen2000curse}.

\begin{table*}
\scriptsize
\centering
\caption{Median coverage and standard deviation after 24 hours for \ourtool variations.}
\begin{tabular}{p{1cm}|r@{\hskip 0.1in}c@{\hskip 0.1in}r|r@{\hskip 0.1in}c@{\hskip 0.1in}r|r@{\hskip 0.1in}c@{\hskip 0.1in}r|r@{\hskip 0.1in}c@{\hskip 0.1in}r|r@{\hskip 0.1in}c@{\hskip 0.1in}r|r@{\hskip 0.1in}c@{\hskip 0.1in}r|r@{\hskip 0.1in}c@{\hskip 0.1in}r|r@{\hskip 0.1in}c@{\hskip 0.1in}r|r@{\hskip 0.1in}c@{\hskip 0.1in}r}
& & \multicolumn{15}{c}{\textbf{\qquad \qquad \qquad \qquad \qquad \qquad \qquad \qquad \qquad \qquad \qquad  \ourtool Variations}} \\
\textbf{Target}  & \multicolumn{3}{c|}{\textbf{original}} & \multicolumn{3}{c}{\textbf{p=2}} & \multicolumn{3}{c}{\textbf{default length}} & \multicolumn{3}{c}{\textbf{weighted} $m_1$} & \multicolumn{3}{c}{\textbf{random}} & \multicolumn{3}{c}{\textbf{t=2}} & \multicolumn{3}{c}{\textbf{t=0.5}} & \multicolumn{3}{c}{\textbf{cross program}} \\

\midrule
proj4    & 7,068 & $\pm$ & 175 & 5,490 & $\pm$ & 399 & 6,993 & $\pm$ & 147 & 6,982 & $\pm$ & 172 & 4,991 & $\pm$ & 346 & 6,974 & $\pm$ & 153 & 7,039 & $\pm$ & 181 & 6,991 & $\pm$ & 145 \\
curl     & 10,912 & $\pm$ & 82  & 10,407 & $\pm$ & 85 & \underline{10,926} & $\pm$ & 89 & 10,876 & $\pm$ & 94 & 10,322 & $\pm$ & 99& 10,860 & $\pm$ & 94 & 10,876 & $\pm$ & 54 & 10,858 & $\pm$ & 103 \\
freetype & 11,491 & $\pm$ & 397 & 10,960 & $\pm$ & 508 & 11,240 & $\pm$ & 429 & 11,204 & $\pm$ & 419 & 10,544 & $\pm$ & 427 & 11,213 & $\pm$ & 366 & 11,366 & $\pm$ & 290 & \underline{11,520} & $\pm$ & 302 \\
bloaty   & 6,028 & $\pm$ & 73 & 5,408 & $\pm$ & 276 & \underline{6,341} & $\pm$ & 79 & 5,971 & $\pm$ & 93 & 5,244 & $\pm$ & 158 & 6,018 & $\pm$ & 106 & 6,015 & $\pm$ & 102 & 5,960 & $\pm$ & 104 \\
php      & 16,689 & $\pm$ & 90 & 16,054 & $\pm$ & 75 & 16,676 & $\pm$ & 68 & 16,651 & $\pm$ & 76 & 16,054 & $\pm$ & 75 & 16,658 & $\pm$ & 59 & 16,593 & $\pm$ & 72 & 16,536 & $\pm$ & 51 \\
libxml   & 19,256 & $\pm$ & 215 & 14,783 & $\pm$ & 481 & 19,166 & $\pm$ & 268 & 19,246 & $\pm$ & 225 & 18,994 & $\pm$ & 1,800 & 19,213 & $\pm$ & 278 & 19,236 & $\pm$ & 177 & 19,054 & $\pm$ & 250 \\
sqlite   & 20,091 & $\pm$ & 254 & 17,028 & $\pm$ & 729 & 19,934 & $\pm$ & 318 & \underline{20,196} & $\pm$ & 691 & 16,675 & $\pm$ & 1,324 & \underline{20,188} & $\pm$ & 851 & \underline{20,176} & $\pm$ & 397 & \underline{20,100} & $\pm$ & 968 \\
libpng   & 2,006 & $\pm$ & 3 & 1,980 & $\pm$ & 19 & 2,005 & $\pm$ & 3 & 2,004 & $\pm$ & 3 & 1,977 & $\pm$ & 16  & 2,006 & $\pm$ & 18 & 2,006 & $\pm$ & 24 & 2,005 & $\pm$ & 24 \\
libpcap  & 2,840 & $\pm$ & 131 & 2,417 & $\pm$ & 122 & 2,830 & $\pm$ & 106 & \underline{2,852} & $\pm$ & 132 & 2,307 & $\pm$ & 114 & \underline{2,841} & $\pm$ & 144 & 2,817 & $\pm$ & 135 & 2,769 & $\pm$ & 105 \\
openssl  & 5,827 & $\pm$ & 5 & 5,821 & $\pm$ & 7 & 5,822 & $\pm$ & 6 & 5,824 & $\pm$ & 7 & 5,822 & $\pm$ & 6 & 5,824 & $\pm$ & 6 & 5,804 & $\pm$ & 16 & 5,819 & $\pm$ & 8 \\
lcms     & 2,004 & $\pm$ & 348 & 1,964 & $\pm$ & 211 & 1,958 & $\pm$ & 218 & \underline{2,010} & $\pm$ & 239 & 1,585 & $\pm$ & 369  & 2,004 & $\pm$ & 223 & 1,988 & $\pm$ & 175 & 2,002 & $\pm$ & 170 \\
re2      & 2,879 & $\pm$ & 5 & 2,836 & $\pm$ & 18 & 2,877 & $\pm$ & 4 & 2,878 & $\pm$ & 5 & 2,860 & $\pm$ & 29 & \underline{2,880} & $\pm$ & 3 & 2,877 & $\pm$ & 4 & 2,878 & $\pm$ & 4 \\
json     & 520 & $\pm$ & 0 & 520 & $\pm$ & 0 & 520 & $\pm$ & 0 & 520 & $\pm$ & 0 &  520 & $\pm$ & 0.2 & 520 & $\pm$ & 0 & 520 & $\pm$ & 0 & 520 & $\pm$ & 0 \\
\bottomrule
\multicolumn{25}{r}{ Variations that perform better than the original are shown \underline{underlined}.} \\
\end{tabular}
\label{tab:results_ablation}
\end{table*}

\subsubsection{Selecting the Length of the Mutator Sequence}
The length $l$ of the mutator sequence $\langle x_1, x_2, ..., x_l \rangle$ is another hyperparameter of \ourtool.
In \Cref{sec:methodology} we described the MAB algorithm \ourtool uses to determine the length $l$, and here we investigate its impact on \ourtool performance.
To do so, we run a variation of \ourtool where $l$ follows the default AFL++ distribution (see \Cref{sec:background}). We call this variation \ourtool{}\textsubscript{default length}.

From \Cref{tab:results_ablation}, we see that \ourtool{}\textsubscript{default length} performs better than or equal to \ourtool in four out of thirteen programs.
For the other nine programs, \ourtool performs better.
Hence, we conclude that the MAB-optimized sequence length contributes to the performance of \ourtool.

\subsubsection{Selecting the First Mutator}
Our proposed algorithm for generating mutator sequences specifies how to select the next mutator $x_{n}$ given the previous mutator in the sequence $x_{n-1}$. 
This leaves open the question of how to select the first mutator $x_1$ of the sequence $\langle x_1, x_2, ..., x_l \rangle$.
We investigate two viable answers:
\begin{itemize}[leftmargin=1em]
    \item a uniformly random selection of $m_1$ (i.e., the default variation), 
    \item a weighted selection of $m_1$ that gives a higher probability to mutators that have produced more interesting inputs.
\end{itemize}
We call the latter \ourtool{}\textsubscript{weighted $m_1$}.
We see that \ourtool performs better than or equal to \ourtool{}\textsubscript{weighted $m_1$} in ten out of thirteen programs.
Hence, we conclude that a uniformly random selection of $m_1$ is preferred over a weighted selection.
This may seem counterintuitive since the weighted selection starts the sequence with a more ``promising'' mutator.
Our interpretation is that the weighted selection makes the first mutator more deterministic, reducing the overall randomness of the mutator sequence, which in turn reduces the fuzzer exploration.

\subsubsection{Using a Random Matrix}
To create a meaningful baseline, we replace the matrix $N(i,j)$ that holds the number of interesting inputs generated by combining mutator $i$ with mutator $j$ with a random matrix. 
We call this resulting baseline \ourtool{}\textsubscript{random} and show its performance in \Cref{tab:results_ablation}.
We see that \ourtool{}\textsubscript{random} performs worse than \ourtool in all target programs, which increases our confidence that the performance of \ourtool is a result of its design and not of randomness.

\subsubsection{Selecting the Training Time}
Previous work on machine learning based fuzzers~\cite{neuzz,mtfuzz} also follows a two-phase process like ours, where the first hours of the fuzzing budget are dedicated for training and the rest of them for inference.
For example, Neuzz~\cite{neuzz} runs AFL for one hour to collect data about which bytes of the seed are more likely to yield an interesting input when mutated.
For this reason, we select $T_{train}=1$ hour as a starting option for our training phase.
To understand the impact of $T_{train}$, we also run two variations with $T_{train}=0.5$ hours and $T_{train}=2$ hours, named \ourtool{}$_{t=0.5}$ and \ourtool{}$_{t=2}$ respectively. 
We see that \ourtool (with $T_{train}=1$) performs at least as good as \ourtool{}$_{t=2}$ in ten out of thirteen target programs and at least as good as \ourtool{}$_{t=0.5}$ in eleven out of thirteen programs.

\subsubsection{Cross-Program Generalization of Learned Probability}
One question that arises is whether the learned probabilities in one target program are transferable to another program.
In other words, whether there is a universal probability of mutation pairs that works well in all programs.
To answer this, we first qualitatively analyze the learned probability of the thirteen programs.
We find that in \emph{nine} programs the learned probability follows a similar pattern as the one shown in \Cref{fig:matrices}.
For the other \emph{four} programs (\texttt{libpng}, \texttt{libpcap}, \texttt{openssl}, and \texttt{lcms}), we see four distinct patterns.
We randomly select one of the nine programs (\texttt{freetype}) and use the probability learned on that program to guide the mutations in the other programs.
The training phase is disabled since no learning is required and the guided mutation phase takes up the whole fuzzing campaign (24 hours);
We call this variation \ourtool{}\textsubscript{cross program}.

From \Cref{tab:results_ablation}, we see that in the four programs that exhibit a different pattern than \texttt{freetype}, the performance of \ourtool{}\textsubscript{cp} is lower than or equal to \ourtool{}\textsubscript{random}.
In the other nine programs, \ourtool{}\textsubscript{cp} performs worse than \ourtool{}\textsubscript{random} only in two programs, while it performs even better than \ourtool in two other programs.
These results indicate that, although cross-program generalization is possible between some (not all) programs, training in each program individually generally leads to better results.

We notice that in \texttt{freetype}, \ourtool{}\textsubscript{cp} performs better than \ourtool.
This makes sense: the training phase, where only two mutators are stacked together, is disabled and the fuzzer remains in the guided mutation phase for 24 hours instead of 23.
This suggests that the training phase slows down \ourtool, making the effectiveness of the guided mutation phase even higher.
To validate, we check the performance of \ourtool after one hour (the end of the training phase) for the ten programs in which \ourtool outperformed the baselines in \Cref{ssec:evaluation_rq2}.
We find that only in four of them \ourtool was better after one hour; in the other six programs, \ourtool started from a disadvantage and surpassed the baselines in the next 23 hours of the guided mutation phase.

The ablation study increases our confidence that \ourtool performance is a result of the effective harness of the interaction effect between mutators and quantifies the effect of hyperparameters.

\section{Discussion}\label{sec:discussion}
In this section, we interpret the outcomes of the training phase of \ourtool (\Cref{ssec:what_our_fuzzer_learns}) and discuss the implications of our findings to future research (\Cref{ssec:future_research}).

\subsection{What does \ourtool learn during training?}\label{ssec:what_our_fuzzer_learns}
The performance of \ourtool motivates a qualitative analysis of the learned probability $\Pr(m_n  \mid m_{n-1})$ to see which mutators (or mutator families) are better combined with which.
For example, the learned probability for the \texttt{freetype} target program is shown in \Cref{fig:matrices} (right).
\emph{Unit} mutators that apply simple, lightweight transformations have a black label font, while \emph{chunk} mutators, that disruptively transform the seed have a blue label font.
We observe the following pattern: 
if the first mutator (row) is a unit mutator, the second mutator (column) is more likely to be a chunk mutator.
On the other hand, if the first mutator is a chunk mutator, the second mutator follows a roughly uniform distribution.
A similar pattern is followed in eight other target programs.
For the target program \texttt{openssl}, however, we observe a different pattern:
unit mutators dominate over chunk mutators.
This means that a mutator sequence sampled from this learned probability will have mostly unit mutators.
This could be because the input has a strict format that easily breaks with chunk mutators.
A similar, but weaker pattern is observed for the \texttt{libpng} target program.
Finally, for the \texttt{libpcap} target program the pattern has nothing to do with unit or chunk mutators.
Specifically, mutators that apply simple arithmetics (IDs \num{7}-\num{18}) tend to work better when stacked on top of each other, and also when followed by clone mutators (IDs \num{29} and \num{30}).

We note that the space of all possible mutator sequences is the same (or a subset, if $N\left(i, j\right)$ contains zeros) for \ourtool and the baselines.
This means that \ourtool does not generate a new sequence that cannot be generated by the baselines, rather it generates the more interesting sequences earlier in the fuzzing campaign. 

\subsection{Implications for Future Research}\label{ssec:future_research}
We mention here the similarity of generating mutator sequences with our proposed method to generating text (token sequences, where tokens can be characters, words, or any other token type) using bigram language models (LMs)~\cite{bigramLMs}.
The goal of a LM is to learn the conditional probability $P\left(x_n \mid x_1x_2...x_{n-1}\right)$ of the token $x_n$ given the previously generated tokens $x_1x_2...x_{n-1}$.
Bigram language models, an early predecessor of modern powerful LMs~\cite{gpt4,llama}, approximate this probability by looking only at the last token $x_{n-1}$.
The training process consists of computing the frequency of all bigrams $\langle i,j \rangle$ appearing in the training set.
The inference of the next character consists of feeding the previously generated character to the probability distribution defined by normalizing the calculated bigram frequencies.
Although the problem domain is different, our two-phase mutation strategy is inspired---up to a certain degree---by the bigram language models.
This similarity yields the question of whether we can apply more advanced text generation techniques to the problem of generating mutator sequences.
For example, to train a Recurrent Neural Network (RNN)~\cite{rnns} or a Transformer~\cite{transformers} on a dataset of successful mutator sequences.
Information about the seed on which the successful mutator was applied could also be encoded as additional context.
These ideas are applicable in light of our results, which show that information about the previous mutator can help select the next mutator.

Another line of future work stems from our finding that selecting the first mutator at random tends to outperform a weighted selection, which we interpret as increased randomness in the generated sequences.
The overall randomness of our mutation strategy can be increased by using a different normalization than simply dividing by the sum in \Cref{eq:1}.
For example, the softmax~\cite{deep_learning} normalization comes with the temperature hyperparameter $T$, where higher $T$ yields less randomness, and lower $T$ yields higher randomness.
Controlling this hyperparameter would be a way to account for the exploration/exploitation trade-off in the mutation strategy.

Finally, the idea of considering the previous state in a mutator sequence can be generalized beyond mutators: it would be interesting to predict the next position to mutate, given the position mutated by the previous mutator in the sequence. 
This work would complement related work that deals with the problem of targeting mutators to specific positions of the seed~\cite{vuzzer,neuzz,fairfuzz}.

\section{Threats to Validity}\label{sec:threats_to_validity}
We present here the threats to the validity of our study.

\noindent\textbf{External Validity.}
The target programs we use in our experiments are open-source C libraries, hence we do not make claims beyond that.
Overfitting a specific set of target programs is a threat to validity in fuzzer evaluation~\cite{sok}.
To mitigate this threat, we systematically reason about the selection of our target programs.
MAGMA comes with nine target programs, which is computationally affordable, so we use all of them.
FuzzBench, on the other hand, comes with 28 target programs, which is beyond our computational budget.
For this reason, we randomly select 13 target programs, by considering the union of the target programs used in two recent related works.
Not comparing to meaningful state-of-the-art fuzzers is another threat, which we mitigate by comparing a) against AFL++, which is the fuzzer we build on top of, and b) against MOPT, which optimizes the selection probability of each mutator in isolation.

\smallskip
\noindent\textbf{Internal Validity.}
The correctness of the implementation of any new tool is a threat to validity, which we mitigate by open-sourcing our code to be assessed by the community~\cite{ours}.
We run statistical tests to increase the confidence that the observed performance difference is not a result of the inherent randomness of fuzzing but of the more effective mutator sequences.
Also, we run an ablation study to increase the confidence that the performance of \ourtool comes from leveraging the interaction effect between mutators.
Finally, measuring fuzzer performance based on proxy measures such as code coverage alone poses a threat to fuzzer evaluation~\cite{sok}.
We mitigate this threat by running experiments on bugs from real-world programs using the MAGMA benchmark.

\section{Conclusion}\label{sec:conclusion}
This work investigates and proposes a method to leverage the interaction effect between mutators in greybox fuzzers.

We investigate the interaction effect by fitting a linear model to a dataset that contains the effectiveness for all possible mutator pairs of AFL-based fuzzers.
We find that the interaction term of the model explains a statistically significant portion of the model's variance, thus it can affect the effectiveness of mutator sequences.

We investigate whether this finding can be leveraged in practice by developing \ourtool, a fuzzer that generates mutator sequences by using information about the previously selected mutator to select the next one.
Our empirical evaluation shows that \ourtool outperforms AFL++ and MOPT in terms of achieved code coverage, and also detects a bug that none of these fuzzers was able to detect.
Given that AFL++ uses a fixed selection probability and MOPT optimizes the selection probability of each mutator in isolation, our results show that mutator strategies that account for the interaction effect can be more effective.

\begin{acks}
K. Kitsios and A. Bacchelli gratefully acknowledge the support of the Swiss National Science Foundation through the SNSF Project 200021\_197227. The authors would also like to thank the Swiss Group for Original and Outside-the-box Software Engineering (CHOOSE) for sponsoring the trip to the conference.
\end{acks}

\bibliographystyle{ACM-Reference-Format}
\bibliography{bibliography}

\end{document}